\documentclass[12pt]{article}
\usepackage{amsfonts}
\usepackage{graphicx}
\usepackage{latexsym,amsmath,color}
\usepackage{natbib}
\usepackage[colorlinks=true,linkcolor=black,urlcolor=black]{hyperref}
\usepackage[english]{babel}
\usepackage{multicol}
\usepackage[top=2.5cm,bottom=3cm,right=2.5cm,left=2.5cm]{geometry}
\hypersetup{citecolor=black}

\usepackage{etoolbox}

\makeatletter

\DeclareRobustCommand\citepos
  {\begingroup\def\NAT@nmfmt##1{{\NAT@up##1's}}%
   \NAT@swafalse\let\NAT@ctype\z@\NAT@partrue
   \@ifstar{\NAT@fulltrue\NAT@citetp}{\NAT@fullfalse\NAT@citetp}}

\pretocmd{\NAT@citex}{%
  \let\NAT@hyper@\NAT@hyper@citex
  \def\NAT@postnote{#2}%
  \setcounter{NAT@total@cites}{0}%
  \setcounter{NAT@count@cites}{0}%
  \forcsvlist{\stepcounter{NAT@total@cites}\@gobble}{#3}}{}{}
\newcounter{NAT@total@cites}
\newcounter{NAT@count@cites}
\def\NAT@postnote{}
\def\NAT@hyper@citex#1{%
  \stepcounter{NAT@count@cites}%
  \hyper@natlinkstart{\@citeb\@extra@b@citeb}#1%
  \ifnumequal{\value{NAT@count@cites}}{\value{NAT@total@cites}}
    {\if*\NAT@postnote*\else\NAT@cmt\NAT@postnote\global\def\NAT@postnote{}\fi}{}%
  \ifNAT@swa\else\if\relax\NAT@date\relax
  \else\NAT@@close\global\let\NAT@nm\@empty\fi\fi
  \hyper@natlinkend}
\renewcommand\hyper@natlinkbreak[2]{#1}

\patchcmd{\NAT@citex}
  {\ifNAT@swa\else\if*#2*\else\NAT@cmt#2\fi
   \if\relax\NAT@date\relax\else\NAT@@close\fi\fi}{}{}{}
\patchcmd{\NAT@citex}
  {\if\relax\NAT@date\relax\NAT@def@citea\else\NAT@def@citea@close\fi}
  {\if\relax\NAT@date\relax\NAT@def@citea\else\NAT@def@citea@space\fi}{}{}
\patchcmd{\NAT@cite}{\if*#3*}{\if*\NAT@postnote*}{}{}\makeatother

\let\oldtheequation\theequation
\makeatletter
\def\tagform@#1{\maketag@@@{\ignorespaces#1\unskip\@@italiccorr}}
\renewcommand{\theequation}{(\oldtheequation)}
\makeatother

\begin{document}

\renewcommand{\tablename}{Table}

\title{A lack-of-fit test for quantile regression models with high-dimensional covariates}
\author{\small{Mercedes Conde-Amboage$^{1}$, C\'esar S\'anchez-Sellero$^{1,2}$  and Wenceslao Gonz\'alez-Manteiga$^{1}$}}
\small{\date{}}
\maketitle

\footnotetext[1]{Department of Statistics and Operations Research. University of Santiago de Compostela (Spain).}
\footnotetext[2]{Corresponding author. e-mail: \href{mailto:cesar.sanchez@usc.es}{cesar.sanchez@usc.es}.}

\begin{abstract}
We propose a new lack-of-fit test for quantile regression models that is suitable even with high-dimensional covariates. The test is based on the cumulative sum of residuals with respect to unidimensional linear projections of the covariates. The test adapts concepts proposed by Escanciano (Econometric Theory, 22, 2006) to cope with many covariates to the test proposed by He and Zhu (Journal of the American Statistical Association, 98, 2003). To approximate the critical values of the test, a wild bootstrap mechanism is used, similar to that proposed by Feng et al. (Biometrika, 98, 2011). An extensive simulation study was undertaken  that shows the good performance of the new test, particularly when the dimension of the covariate is high. The test can also be applied and performs well under heteroscedastic regression models. The test is illustrated with real data about the economic growth of 161 countries.


\end{abstract}

\section{Introduction}

Let us consider a regression setting where a quantile of the response variable of interest, $Y$, is expressed as a function of a vector of explanatory variables, $X$. The resulting regression model can then be denoted by
$$Y=g(X)+\varepsilon$$
where $g$ represents the quantile regression function; and the error, $\varepsilon$, has a conditional $\tau$-quantile equal to zero, $P(\varepsilon <0|X=x)=\tau$ for almost all $x$.

Quantile regression models have been receiving increased attention in the literature, due to their flexibility for general error distributions and because they provide a more detailed description of the conditional distribution of the response, compared to classical mean regression. \citet{KyB78} can be considered as the seminal work on the estimation of linear quantile regression models. The main concept is to exploit that the $\tau$-quantile, $g$, of a variable minimizes the expectation,
$$E\left(\rho_{\tau}(Y-g)\right),$$
where $\rho_\tau(r)=\tau r I(r> 0)+(\tau-1)r I(r<0)$, and $I(\cdot)$ denotes the indicator function of an event. Estimation of quantile models is obtained by minimizing the sum of penalized residuals, similarly to the sum of squares in the case of mean regression. That is, given a sample of independent observations, $(X_1,Y_1),\ldots,(X_n,Y_n)$, the coefficients of a linear model, $g(x)=x'\theta$ ($x'$ denotes the transpose of $x$), are estimated as the minimizers of
$$\sum_{i=1}^n \rho_{\tau}\left(Y_i-X_i'\theta\right).$$

The same criterion can be applied to estimate general parametric models, where the regression function is of the type $g(\cdot,\theta)$ and $\theta$ is a parameter to be estimated, and even to nonparametric estimation of the quantile regression function. See \citet{Koen05} for a complete review on quantile regression methods.

We focus on the problem of testing a parametric model of quantile regression. That is, a test of the null hypothesis
$$H_{0} : g \in \mathcal{M}_{\theta}=\{g(\cdot,\theta) \; : \; \theta \in \Theta \subset \mathbb{R}^{q}\}$$
versus a nonparametric alternative.

This problem was addressed by \citet{HyZ03}, who based their test on the process
$$n^{-1/2}\sum_{i=1}^n \psi\left(Y_i-g(X_i,\widehat\theta)\right) \dot{g}(X_i,\widehat\theta) I(X_i\leq t)$$
where $\psi(r)=\tau I(r>0)+(\tau-1)I(r<0)$ is the derivative of $\rho_\tau$, $\dot{g}(x,\theta)=\frac{\partial}{\partial \theta} \; g(x,\theta)$, and $\widehat\theta$ is an estimator of $\theta$. This is an extension to the quantile regression setting of the cumsum process considered by \citet{Stu97} in the mean regression setting. \citet{Zheng98} proposed a U-statistic of the quantities $\psi(Y_i-g(X_i,\theta))$ with smoothing kernel weights, thereby extending the test of \citet{Zheng96} to quantile models. Other specification tests for quantile regression models can be found in \citet{HyS02}, \citet{Wha06}, \citet{Ots08}, \citet{EyV10}, and \citet{EyG14}, among others.

It is well-known that a high (or even moderate) dimension of the covariate can affect the performance of the specification tests. This problem has been addressed by several authors in the mean regression setting, where modified tests have been proposed with better properties for multiple covariates. In particular, \citet{Esc06} applied a cumsum test to one-dimensional projections of the covariates, \citet{LyP08} considered similar one-dimensional projections for a Zheng type test, and \citet{SXZ08} based their test on the residual empirical process marked by proper functions of the regressors.

Little can be found in the literature for lack-of-fit testing adapted to multidimensional covariates in the framework of quantile regression. \citet{Wil08} used a He and Zhu type test and defined some ranks over the covariate. This proposal has the virtue of simplicity but does not provide an omnibus test, i.e., it is not consistent for all alternatives.

We propose and study a lack-of-fit test for parametric models of quantile regression, with good properties for multidimensional covariates and consistent for all alternatives. In Section 2 we present the new  He and Zhu type test calculated on one-dimensional projections of the covariates. A bootstrap method is also proposed to approximate the critical values of the test. Section 3 contains a simulation study where the performance of the test is studied under homo- and heteroscedastic models, with different error distributions and with increasing dimension of the covariate. We compare the proposed test with a He and Zhu test. In Section 4 the test is applied to real data, and we provide some concluding remarks and extensions in Section 5.

\section{The proposed method}\label{section1}

\subsection{The test}

The strategy to improve the performance of the test with multiple covariates consists of applying a lack-of-fit test to one-dimensional projections of the covariates. This is motivated by a fundamental result, that states that the null hypothesis, $H_{0} : g \in \mathcal{M}_{\theta}$, holds if and only if, for some $\theta_{0}\in \Theta \subset \mathbb{R}^{p}$, and for any $\beta \in \mathbb{R}^{d}$ with $\|\beta\|=1$,
$$P[Y-g(X,\theta_{0})\leq 0 \; | \; \beta'X]=\tau$$
almost surely. This is an immediate extension of Lemma 1 in \citet{Esc06} to the quantile regression setting.

If the true parameter $\theta_0$ was known, the test could be based on the process
$$R_{n}(\beta,u)= n^{-1/2}\sum_{i=1}^{n} \psi\left(Y_i-g(X_i,\theta_{0})\right)
\dot{g}\left(X_{i},\theta_{0}\right)I\left(\beta'X_{i}\leq u\right).$$

Otherwise, an estimator $\widehat{\theta}$ is substituted, yielding the process useful for lack-of-fit testing of the parametric model
$$R_{n}^{1}(\beta,u)= n^{-1/2}\sum_{i=1}^{n} \psi\left(Y_i-g\left(X_i,\widehat\theta\right)\right)
\dot{g}\left(X_{i},\widehat{\theta}\right)I\left(\beta'X_{i}\leq u\right) .$$

The test statistic is then defined as
\begin{equation} \label{eq:tn}
T_{n}=\mbox{largest eigenvalue of } \; \int_{\Pi} R_{n}^{1}(\beta,u)[R_{n}^{1}(\beta,u)]'F_{n,\beta}(du)d\beta ,
\end{equation}
where $\Pi=\mathbb{S}_{d} \times [-\infty,+\infty]$, $\, \mathbb{S}_{d}$ is the unit sphere on $\mathbb{R}^{d}$, and $F_{n,\beta}$ is the empirical distribution of the projected covariates $\beta'X_1,\ldots,\beta'X_n$.

The process $R_n^1$ is similar to that proposed by \citet{Esc06}, with two differences: the loss function is now the quantile loss function, and the gradient vector $\dot{g}(X_{i},\widehat{\theta})$ is introduced following the suggestion of \citet{HyZ03}.

The limit distribution of $R_n$ under the simple null hypothesis, $H_0: g=g(x,\theta_0)$ with $\theta_0$ known, can be expressed as 
$$R_n\stackrel{d}{\rightarrow} R_\infty ,$$
where $R_\infty$ is a Gaussian process with mean zero and covariance given by
$$K(x_1,x_2)=\tau(1-\tau)E\left[\dot{g}(X,\theta_0)\dot{g}^\prime(X,\theta_0)
  I(\beta_1'X\leq u_1) I(\beta_2'X\leq u_2)\right],$$
where $x_1=(\beta_1',u_1)'$ and $x_2=(\beta_2',u_2)'$. This result can be obtained similarly to \citet{Esc06}, where the tightness comes from the fact that the family of functions in the definition of $R_n$ is a VC-class of functions.

Under the composite null hypothesis of a parametric model, $H_{0} : g \in \mathcal{M}_{\theta}$, and under certain regularity conditions, the following representation can be obtained:
$$R_{n}^1(\beta,u) = n^{-1/2} \sum_{i=1}^{n}\psi(\varepsilon_i) \left[
I(\beta'X_i\leq u) -S(\beta,u)S^{-1}\right]\dot{g}(X_i,\theta_0)+o_{p}(1)
$$
uniformly in $(\beta,u)$, where $\varepsilon_i=Y_i-g(X_i,\theta_0)$, $i=1,\ldots,n$ are the errors, $f(0|X)$ denotes the conditional density of the error at zero, and the matrices $S$ and $S(\beta,u)$ are defined by
\begin{eqnarray*}
S&=&E[f(0|X)\dot{g}(X,\theta_0)\dot{g}'(X,\theta_0)] \\
S(\beta,u)&=&E[f(0|X)\dot{g}(X,\theta_0)\dot{g}'(X,\theta_0)I(\beta'X\leq u)] .
\end{eqnarray*}

The proof and the subsequent consequences are a combination of arguments given in \citet{HyZ03} and \citet{Esc06}. The representation itself is different from that of \citet{HyZ03}, because we do not assume homoscedasticity. From this representation, the limit distribution of the test statistic, $T_n$, under the null hypothesis can be derived.

Under the alternative, the representation is similar to the previous case, but a new term appears which will be crucial to prove the consistency of the test. Let us assume that the data come from
$$Y_i=g(X_i,\theta_0)+n^{-1/2} h(X_i)+\varepsilon_i\qquad i\in\{1,\ldots,n\},$$
where $\varepsilon_1,\ldots,\varepsilon_n$ are independent errors with conditional $\tau-$quantile equal to zero. The errors are not assumed to be identically distributed. In particular, their density at zero may depend on $X$. With this type of data drawn from the alternative hypothesis, the process allows the following representation:
\begin{eqnarray*}
R_{n}^1(\beta,u) &= & n^{-1/2} \sum_{i=1}^{n}\psi(\varepsilon_i) \left[
I(\beta'X_i\leq u) -S(\beta,u)S^{-1}\right]\dot{g}(X_i,\theta_0) \\
&& + E\left[ f(0|X)h(X)\dot{g}'(X,\theta_0)I(\beta'X\leq u)\right] \\
&& -S(\beta,u)S^{-1}E\left[ f(0|X)h(X)\dot{g}'(X,\theta_0)\right]+o_{p}(1)
\end{eqnarray*}
uniformly in $(\beta,u)$. The second and third summands of the right-hand side are constants reflecting the deviation from the null hypothesis. If the data come from
$$Y_i=g(X_i,\theta_0)+c_n n^{-1/2} h(X_i)+\varepsilon_i\qquad i\in\{1,\ldots,n\},$$
where $c_n$ is a sequence of real numbers converging to infinity (at any rate), then the test statistic, $T_n$, will converge to infinity and the power of the test will converge to one. To obtain this consistency, it is assumed that the sequence $g(x,\theta_0)+c_n n^{-1/2} h(x)$ does not coincide with any element of the parametric model, $\mathcal{M}_{\theta}=\{g(\cdot,\theta) \; : \; \theta \in \Theta \subset \mathbb{R}^{q}\}$, and that $\mbox{Var}(f(0|X)h(X)\dot{g}'(X,\theta))>0$ for any $\theta$.

\subsection{Bootstrap approximation}

The approximation of critical values is a crucial issue in lack-of-fit testing. One possible solution would be to use the limit distribution. However, this would require an estimate of the limit variance which involves the estimation of complicated unknown quantities. Furthermore, the convergence to the limit distribution could be slow. Another possibility could be to use the representations as given above. Then, a bootstrap method based on multipliers can be considered (see \citet{HyZ03}). The approximation by a multipliers bootstrap is generally better than the limit distribution, but still requires estimating many unknown quantities. \citet{HyZ03} assume homoscedasticity, so the conditional density of the error at zero, $f(0|X)$, does not have to be estimated. On the other hand, \citet{EyG14} allow for heteroscedasticity and use a multipliers bootstrap, which requires an estimate of the conditional density $f(0|X)$ by a smoothing method.

We propose a bootstrap approximation based on drawing new bootstrap samples, $(X_1,Y_1^{\star}),\ldots,$ \newline $(X_n,Y_n^{\star})$, where
$$Y_{i}^{\star}=g(X_{i},\widehat{\theta})+\varepsilon_{i}^{\star}\qquad i=1,\ldots,n.$$
$\widehat\theta$ is the parameter estimate obtained from the original sample, and $\varepsilon_i^{\star}=w_{i}|r_{i}|$, where $r_i=Y_i-g(X_i,\widehat\theta)$ are the residuals from the original sample. The multipliers, $w_i$, are independently generated from a common distribution with $\tau$-quantile equal to zero. Following the proposal by \citet{FHH11}, the absolute values of the residuals are used to construct the bootstrap errors, which is a convenient modification of wild bootstrap for quantile regression. Regarding the multipliers distribution, we adopt the two-point distribution with probabilities $(1-\tau)$ and $\tau$ at $2(1-\tau)$ and $-2\tau$, respectively, that was proposed by \citet{FHH11} to satisfy their Conditions 3, 4 and 5. Note that other common multipliers distributions for mean regression, generally with the only condition that the variance is one and occasionally with the condition that the third moment is one (see \citet{M93} for a two-point multipliers distribution in the mean regression), do not satisfy Conditions 4 and 5 required by \citet{FHH11} to establish consistency of the bootstrap for quantile regression.

The advantage of the proposed bootstrap approximation for the lack-of-fit test, in comparison to existing methods such as those proposed by \citet{HyZ03} and \citet{EyG14}, is that it allows consideration of heteroscedastic regression models of any type without needing to estimate complicated quantities in the representations, and in particular without estimating the conditional density $f(0|X)$ by smoothing methods.

Once the bootstrap sample is generated, the test statistic is computed in the same way as for the original sample, obtaining $T_n^*$. If a number, $B$, of bootstrap samples are generated, then $T_{n,1}^{\star},\ldots,T_{n,B}^{\star}$ represents the $B$ bootstrap values of the test statistic. The p-value of the test may be approximated by the proportion of bootstrap values not smaller than the original test statistic, i.e., $(1/B)\sum_{b=1}^B I(T_n\leq T_{n,b}^{\star})$.

The validity of this bootstrap mechanism comes from the representation of the process $R_n^1$ under the null hypothesis, in terms of the true errors plus the parameters estimation,
$$R_{n}^1(\beta,u) = n^{-1/2} \sum_{i=1}^{n}\psi(\varepsilon_i) \dot{g}(X_i,\theta_0) I(\beta'X_i\leq u) -S(\beta,u) \sqrt{n} \left(\widehat{\theta}-\theta\right)+o_{p}(1)
$$
uniformly in $(\beta,u)$. A similar representation can be derived for the bootstrap process conditionally on the original sample, where the convergence of the bootstrap version of the estimation error, $\sqrt{n} (\widehat{\theta}^*-\widehat{\theta})$, was established in Theorem 1 of \citet{FHH11}.

\subsection{Computational aspects}

Tests that face the curse of dimension usually require additional algorithms over other more common model checks. In particular, \citet{Esc06} and \citet{SXZ08} are based on \citet{Stu97}'s test and require additional computations over this original method. Similarly, \citet{LyP08} is a test for high-dimensional covariates that is based on \citet{Zheng96}'s test, and requires an optimization algorithm over a set of Zheng-type statistics. The proposed method here is an adaptation of \citet{HyZ03}'s test to high-dimensional covariates with a procedure similar to that given by \citet{Esc06}. One important virtue of this procedure is the ease of computation and that the amount of computations does not grow dramatically with the dimension of the covariate.
To illustrate this, recall that our test statistic, $T_n$, was defined in \ref{eq:tn} as the largest eigenvalue of a Cramer-von-Mises norm of the process $R_n^1$. Following \citet{Esc06}, one can show that $T_n$ can be expressed as
$$T_{n}= \mbox{largest eigenvalue of} \; n^{-2} \; \sum_{i=1}^{n} \sum_{j=1}^{n} \psi(r_{i})\psi(r_{j})\dot{g}(X_{i},\widehat{\theta}) \dot{g}'(X_{j},\widehat{\theta}) A_{ij\bullet},$$
where $A_{ij\bullet}$ is given by
$$A_{ij\bullet}=\sum_{r=1}^{n} A_{ijr}^{(0)} \; \frac{\pi^{d/2 - 1}}{\Gamma\biggl(\frac{d}{2}+1\biggr)},$$
where $A_{ijr}^{(0)}$ is the complementary angle between the vectors $(X_{i}-X_{r})$ and $(X_{j}-X_{r})$ measured in radians, $\Gamma$ is the gamma function, and $d$ is the dimension of the covariate, $X$. Thus, the total number of computations required to obtain the test statistic depends on the dimension, $d$, only at a linear rate, which is the same rate required by \citet{HyZ03}'s test, and much less than the optimization in $d$ dimensions required by other methods in the literature. Note also that the matrix $A_{ij\bullet}$, which is the most expensive in computation time, does not need to be computed for each bootstrap sample. All these computational properties are particularly useful in the case of high-dimensional or functional covariates, see \citet{GP13} for an illustration in the mean regression functional context.

Table \ref{table0} shows the mean of the times required by $1000$ original samples with $B=500$ bootstrap replications, in units of seconds per original sample. The data are drawn from Model 8, whose details are given in the next section, and the sample size is $n=100$. The dimension of the covariate is $d=t+2$. As expected, the new test requires more computations than \citet{HyZ03}'s test, but the differences are quite small, and the amount of computations does not dramatically grow with the dimension, even for very large dimensions. The gain of power from the new test, shown in the next section, justifies the small increase in the computation time.

\begin{table}
\begin{center}
\scalebox{0.65}[1]{
\begin{tabular}{|c|c c c c c c c c|}
\cline{2-9}
\multicolumn{1}{c|}{}& $t=0$ & $t=2$ & $t=6$ & $t=10$ & $t=20$ & $t=30$ & $t=40$ & $t=50$ \\
\hline
\hline
Proposed test & 2.76 & 2.84 & 2.85 & 2.91 & 2.91 & 3.10 & 3.20 & 3.38 \\
HZ test      & 2.71 & 2.51 & 2.81 & 2.56 & 2.92 & 2.83 & 2.85 & 2.77 \\
\hline
\end{tabular}}
\caption{Computational times (seconds per sample) associated with our proposed lack-of fit test (Proposed test) and with the test proposed by \citet{HyZ03} (HZ test) as a function of the dimension $t+2$ of the covariate.}
\label{table0}
\end{center}
\end{table}

\section{Simulation study} \label{sec3}

We study the performance of our proposed method under the null and the alternative hypotheses using a Monte Carlo simulation. In all experiments, the number of simulated original samples was $1000$, the number of bootstrap replications $B=500$, and the multipliers for the bootstrap approximation followed the two-point distribution given in Section 2.2.

We first focus on the behavior under the null hypothesis, to check the adjustment of the significance level. We simulate values for the following quantile regression models with $\tau=0.5$:
\begin{equation*}
\begin{split}
& \mbox{Model 1:} \; \; \; Y=1+X_{1}+X_{2}+\varepsilon ,\\
& \mbox{Model 2:} \; \; \; Y=1+X_{1}+X_{2}+X_{3}+X_{4}+X_{5}+\varepsilon , \\
& \mbox{Model 3:} \; \; \; Y=1+X_{1}+X_{2}+f(X_{1})\varepsilon ,\\
\end{split}
\end{equation*}
where $X_{i} \in \mbox{Uniform}(0,1)$ for $i=\{1,\cdots,5\}$, and they are mutually independent; and $f(x)=x+0.5$ and $\varepsilon \in N(0,1)$ is the unknown error, which is drawn independently of the covariates. In Models 1 and 3 the null hypothesis is the linear model in $X_1$ and $X_2$ versus an alternative that includes any dependence of $Y$ on $X_1$ and $X_2$. In Model 2 the null hypothesis is the linear model in the five explanatory variables versus any dependence on them. Model 1 represents a common homoscedastic model with small dimension of the covariate. Model 2 is intended to show the possible effect of a larger dimension on the level. Model 3 is useful to show the possible effect of heteroscedasticity on the level.

Table \ref{table1} shows the proportions of rejections associated with different sample sizes,  $n$, and for different nominal significance levels, $\alpha$. The proposed test works well in a homoscedastic context (Models 1 and 2) as well as in a heteroscedastic context (Model 3) even for small sample sizes. Comparing Models 1 and 2, the increase of the dimension of the explanatory variables does not have a negative impact on the adjustment of the significance level of the test. These are important, because our bootstrap mechanism was designed to work under heteroscedastic models and the aim of the test itself was to be applied for larger dimensions of the covariate.

\begin{table}
\begin{center}
\scalebox{0.7}[1]{
\begin{tabular}{ c|c c c|c c c|c c c|}
\cline{2-10}
&\multicolumn{3}{c|}{Model 1}&\multicolumn{3}{c|}{Model 2}&\multicolumn{3}{c|}{Model 3}\\
\cline{2-10}
&$\alpha=0.10$&$\alpha=0.05$&$\alpha=0.01$&$\alpha=0.10$&$\alpha=0.05$&$\alpha=0.01$&$\alpha=0.10$&$\alpha=0.05$&$\alpha=0.01$\\
\hline
\hline
\multicolumn{1}{|c|}{n=25} & 0.096 & 0.049 & 0.002 & 0.119 & 0.066 & 0.017 & 0.107 & 0.061 & 0.014 \\
\multicolumn{1}{|c|}{n=50} & 0.112 & 0.047 & 0.008 & 0.112 & 0.053 & 0.014 & 0.099 & 0.045 & 0.005 \\
\multicolumn{1}{|c|}{n=100}& 0.102 & 0.058 & 0.016 & 0.094 & 0.047 & 0.011 & 0.107 & 0.049 & 0.010 \\
\multicolumn{1}{|c|}{n=150}& 0.089 & 0.048 & 0.007 & 0.104 & 0.056 & 0.014 & 0.096 & 0.055 & 0.015 \\
\multicolumn{1}{|c|}{n=200}& 0.100 & 0.048 & 0.010 & 0.106 & 0.049 & 0.010 & 0.100 & 0.054 & 0.015 \\
\hline
\end{tabular}}
\caption{Proportions of rejections associated with our proposed lack-of-fit test for Models 1, 2 and 3.}
\label{table1}
\end{center}
\end{table}

Table \ref{table2} provides the same proportions of rejections for different error distributions and quantiles, restricted to Model 1 and nominal level $\alpha=0.05$. The error distributions are centered standard normal, centered log-normal, and centered exponential with expectation one. That is, $\varepsilon=Z-z_\tau$, where $Z$ follows a standard normal, log-normal, and exponential with expectation one, respectively, and $z_\tau$ is the $\tau$-quantile of the $Z$-distribution. The nominal level is respected under the null hypothesis for all the error distributions considered and orders of the quantile.

\begin{table}
\begin{center}
\scalebox{0.65}[1]{
\begin{tabular}{|cc|c c c c c|}
\cline{3-7}
\multicolumn{2}{c|}{}& $\tau=0.10$ & $\tau=0.25$ & $\tau=0.50$ & $\tau=0.75$ & $\tau=0.90$ \\
\hline
\hline
$\varepsilon\in$ Centered Standard Normal & $n=50 $ & 0.048 & 0.061 & 0.047 & 0.063 & 0.043 \\
               & $n=150$ & 0.052 & 0.053 & 0.048 & 0.057 & 0.051 \\
\hline
$\varepsilon \in$ Centered Log-Normal & $n=50 $ & 0.057 & 0.053 & 0.042 & 0.055 & 0.057 \\
                & $n=150$ & 0.041 & 0.053 & 0.052 & 0.051 & 0.057 \\
\hline
$\varepsilon\in$ Centered Exponential & $n=50 $ & 0.058 & 0.054 & 0.048 & 0.057 & 0.053 \\
                & $n=150$ & 0.060 & 0.056 & 0.059 & 0.049 & 0.046 \\
\hline
\end{tabular}}
\caption{Proportions of rejections associated with our lack-of-fit test for Model 1, for different error distributions and different quantiles, with nominal level $\alpha=0.05$.}
\label{table2}
\end{center}
\end{table}

We now study the performance of the new test under the alternative. To this end, the new test will be compared with that of \citet{HyZ03}. Before doing so, we must remember that \citet{HyZ03} suggested a bootstrap calibration of their test based on an asymptotic representation of the empirical process in a homoscedastic scene. We will verify if this manner of calibrating the test allows a good fit to the significance level for heteroscedastic models. We simulate values of the following regression model with $\tau=0.5$ under the null hypothesis of linearity:
\begin{equation*}
\mbox{Model 4:} \; \; \; Y=1+X_{1}+f(X_{1})\varepsilon ,
\end{equation*}
where $X_{1} \in \mbox{Uniform}(0,1)$, $f(x)=x+0.5$, $\varepsilon \in N(0,1)$, and $X_1$ and $\varepsilon$ are independent.

The proportions of rejections associated with the test proposed by \citet{HyZ03} are shown in Table \ref{table3} for different sample sizes and nominal significance levels. The bootstrap method proposed by \citet{HyZ03} does not work well in a heteroscedastic context. This is due to their representation being only valid under homoscedasticity. However, the proposed bootstrap (Section 2.2) works well for their test also under heteroscedasticity. Therefore, with the aim to make a fair comparison between our proposal and \citet{HyZ03}'s test, subsequently we use a wild bootstrap as given in Section 2.2 to calibrate both lack-of-fit tests.

\begin{table}
\begin{center}
\scalebox{0.65}[1]{
\begin{tabular}{c|c c c | c c c |}
\cline{2-7}
&\multicolumn{3}{c|}{ Wild bootstrap}&\multicolumn{3}{c|}{\small{Bootstrap proposed} }\\
&\multicolumn{3}{c|}{ of Section 2.2}&\multicolumn{3}{c|}{\small{in \citet{HyZ03}}}\\
\cline{2-7}
&$\alpha=0.10$&$\alpha=0.05$&$\alpha=0.01$&$\alpha=0.10$&$\alpha=0.05$&$\alpha=0.01$\\
\hline
\hline
\multicolumn{1}{|c|}{$n=25$}  & 0.103 & 0.057 & 0.014 & 0.441 & 0.305 & 0.142 \\
\multicolumn{1}{|c|}{$n=50$}  & 0.116 & 0.064 & 0.015 & 0.263 & 0.164 & 0.067 \\
\multicolumn{1}{|c|}{$n=100$} & 0.094 & 0.051 & 0.013 & 0.167 & 0.092 & 0.033 \\
\multicolumn{1}{|c|}{$n=150$} & 0.104 & 0.051 & 0.010 & 0.155 & 0.085 & 0.025 \\
\multicolumn{1}{|c|}{$n=200$} & 0.103 & 0.051 & 0.014 & 0.136 & 0.080 & 0.026 \\
\hline
\end{tabular}}
\caption{Proportions of rejections associated with the test proposed by \citet{HyZ03} for the heteroscedastic Model 4 with two types of bootstrap approximations.}
\label{table3}
\end{center}
\end{table}

Once the adjustment of the level of both lack-of-fit tests has been studied, we analyze their performance under the alternative hypothesis. Consider the following regression model associated with quantiles of different orders, $\tau$:
\begin{equation*}
\mbox{Model 5:} \; \; \; Y=1+\frac{1}{5}\left(X_1-X_2 \right)+\varepsilon_{\tau},
\label{modelo5}
\end{equation*}
where $X_{1}, X_{2} \in N(0,1)$ and they are independent, and $\varepsilon=Z-z_{\tau}$,  where $z_{\tau}$ is the $\tau$-quantile of the variable $Z$. $Z$ is drawn independently of $X_1$ and $X_2$. Three possibilities are considered for the distribution of $Z$: standard normal, uniform on the interval $(-1,1)$, and chi-squared with four degrees of freedom.

Table \ref{table4} shows the proportions of rejections for several quantiles and the three error distributions, when the tests are applied to check the no-effect model, i.e., to check the null hypothesis that the quantile regression function is a constant not depending on the covariates. The sample size is fixed to $n=100$. We consider a relatively simple hypothesis and a simple deviation under the alternative, to facilitate the comparison between quantiles of different orders, and to evaluate the effect of the error distribution.

The proposed test is more powerful than \citet{HyZ03}'s test for any of the quantiles and for the three error distributions. The power of the proposed test is symmetric with respect to the order of the quantile around $0.5$ for the symmetric error distributions, which are the standard normal and the uniform in Table \ref{table4}. For the standard normal error distribution, the proposed test is more powerful for the central quantiles (around $0.5$), which can be explained by the higher density at these quantiles. For the uniform error distribution, the density is constant with respect to the quantile, while the factor $\tau(1-\tau)$ appearing in the asymptotic distribution of the proposed test makes the test more powerful for the external quantiles (with orders close to $0$ or $1$). For the chi-squared error distribution, the proposed test is more powerful for the quantiles with smaller order, since the error distribution is asymmetric with higher density at these quantiles.

\begin{table}
\begin{center}
\scalebox{0.7}[1]{
\begin{tabular}{|c c|c c c|c c c|}
\cline{3-8}
\multicolumn{2}{c|}{}&\multicolumn{3}{c|}{Proposed test}&\multicolumn{3}{c|}{HZ test}\\
\cline{3-8}
\multicolumn{2}{c|}{}&$\alpha=0.10$&$\alpha=0.05$&$\alpha=0.01$&$\alpha=0.10$&$\alpha=0.05$&$\alpha=0.01$\\
\hline
\hline
$Z \in N(0,1)$  & $\tau=0.10$ & 0.346 & 0.229 & 0.092 & 0.183 & 0.094 & 0.023 \\
                 & $\tau=0.25$ & 0.498 & 0.362 & 0.180 & 0.210 & 0.121 & 0.030 \\
                 & $\tau=0.50$ & 0.575 & 0.444 & 0.231 & 0.208 & 0.110 & 0.032 \\
                 & $\tau=0.75$ & 0.487 & 0.377 & 0.200 & 0.191 & 0.096 & 0.016 \\
                 & $\tau=0.90$ & 0.357 & 0.245 & 0.102 & 0.128 & 0.052 & 0.007 \\
\hline
$Z \in \mbox{Uniform}(-1,1)$ & $\tau=0.10$ & 0.930 & 0.885 & 0.707 & 0.524 & 0.335 & 0.112 \\
                 & $\tau=0.25$ & 0.866 & 0.789 & 0.593 & 0.397 & 0.242 & 0.066 \\
                 & $\tau=0.50$ & 0.809 & 0.691 & 0.475 & 0.325 & 0.186 & 0.056 \\
                 & $\tau=0.75$ & 0.877 & 0.795 & 0.587 & 0.381 & 0.229 & 0.054 \\
                 & $\tau=0.90$ & 0.945 & 0.872 & 0.693 & 0.382 & 0.193 & 0.027 \\
\hline
$Z \in \chi_{4}^{2}$ & $\tau=0.10$ & 0.315 & 0.207 & 0.076 & 0.144 & 0.078 & 0.018 \\
                      & $\tau=0.25$ & 0.245 & 0.144 & 0.045 & 0.124 & 0.056 & 0.015 \\
                      & $\tau=0.50$ & 0.208 & 0.124 & 0.041 & 0.112 & 0.058 & 0.012 \\
                      & $\tau=0.75$ & 0.141 & 0.070 & 0.022 & 0.115 & 0.058 & 0.017 \\
                      & $\tau=0.90$ & 0.137 & 0.077 & 0.028 & 0.120 & 0.064 & 0.015 \\
\hline
\end{tabular}}
\caption{Proportions of rejections associated with our proposed lack-of-fit test (Proposed test) and with the test proposed by \citet{HyZ03} (HZ test) for Model 5.}
\label{table4}
\end{center}
\end{table}

We now consider a linear model under the null hypothesis and a quadratic deviation under the alternative. The deviation is multiplied by a value $c>0$, to evaluate the effect of the deviation on the power of the test.

\begin{equation*}
\mbox{Model 6:} \; \; \; Y=1+X_1+X_2+c\left(X_1^2+X_2^2+X_1X_2\right) +\varepsilon_{\tau},
\label{modelo6}
\end{equation*}
where $X_{1} \in \mbox{Uniform}(0,1)$, $X_{2} \in N(0,1)$; and $\varepsilon_{\tau}$ is a log-normal distribution centered to the quantile $\tau$, i.e., $\varepsilon_{\tau}=e^{Z}-e^{z_{\tau}}$, where $Z \in N(0,1)$ and $z_{\tau}$ are the $\tau$-quantile of the variable $Z$; and $X_1$, $X_2$ and $\varepsilon_{\tau}$ are drawn independently.

\begin{figure}
\begin{minipage}{.49\linewidth}
\centering
\includegraphics[scale=0.41]{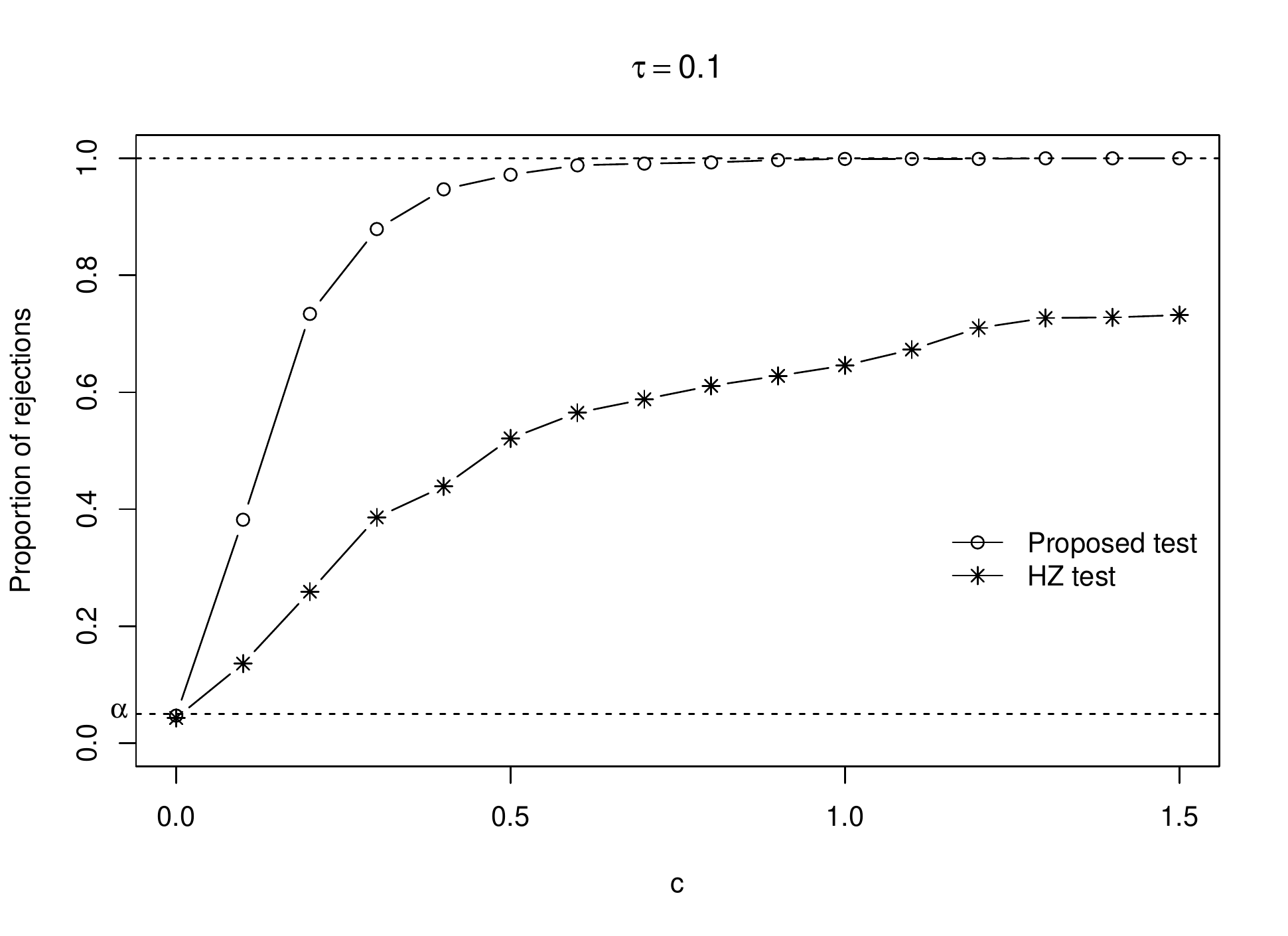}
\end{minipage}
\begin{minipage}{.49\linewidth}
\centering
\includegraphics[scale=0.41]{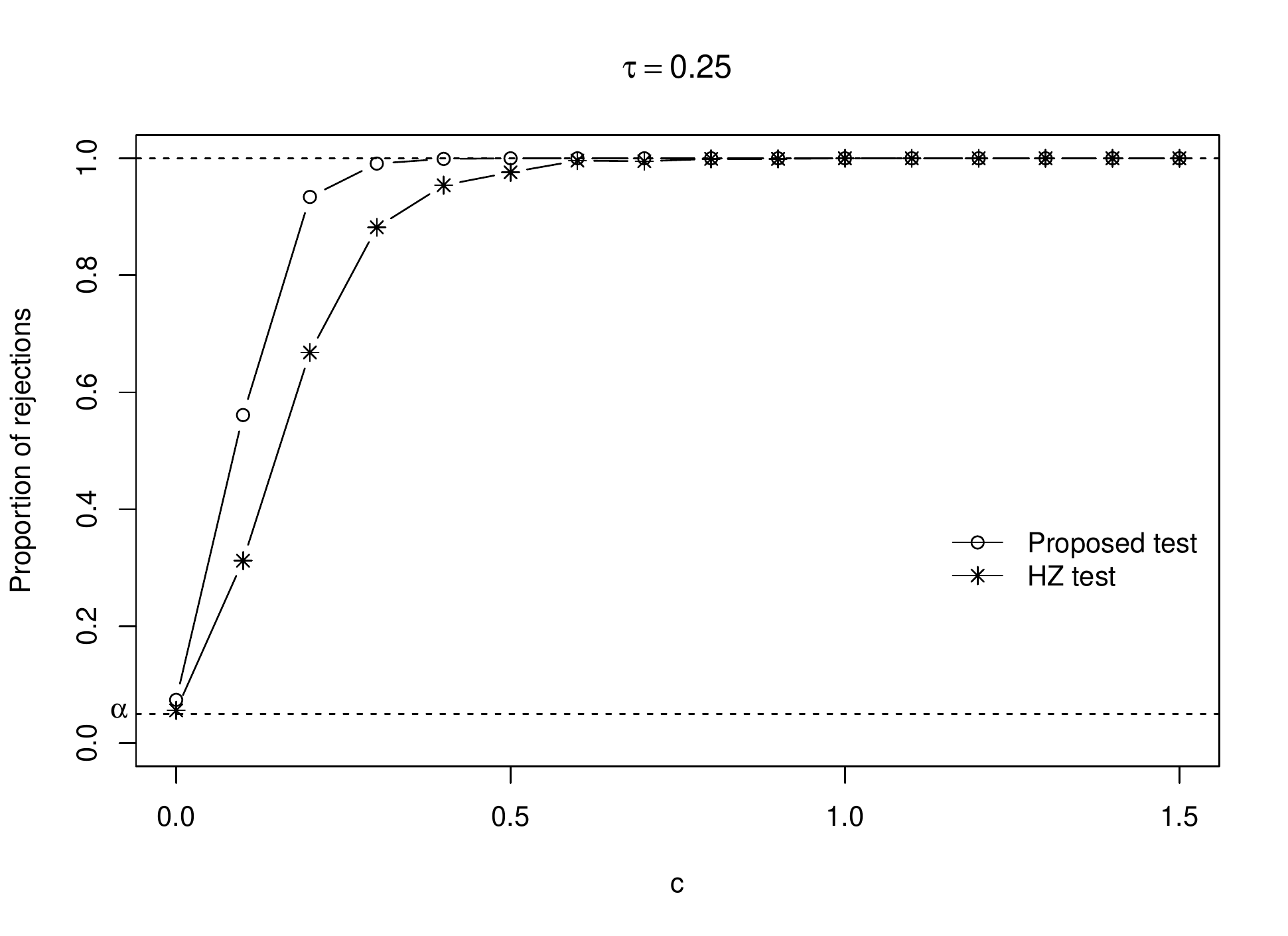}
\end{minipage}
\begin{minipage}{1.0\linewidth}
\centering
\includegraphics[scale=0.41]{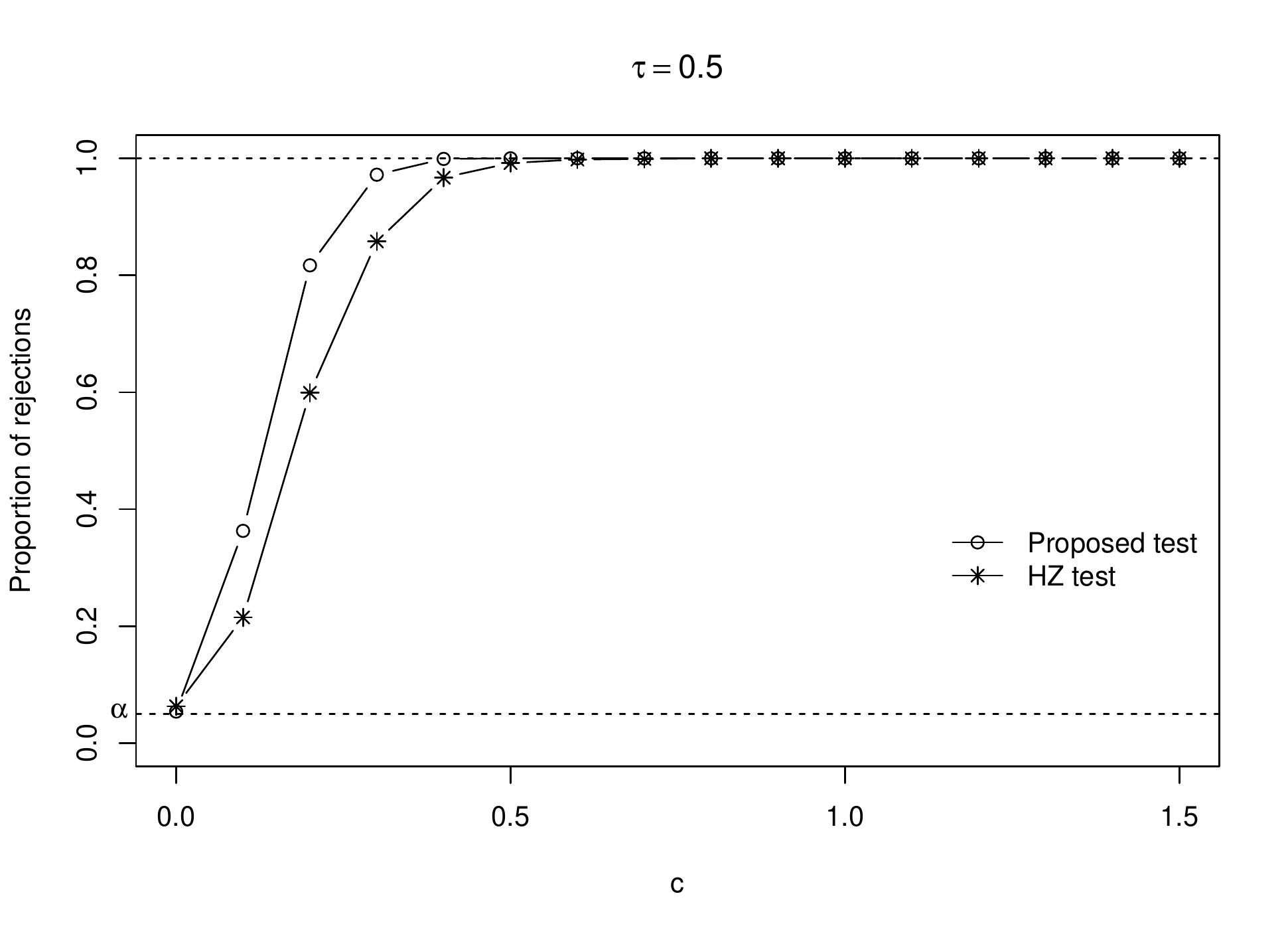}
\end{minipage}
\begin{minipage}{.49\linewidth}
\centering
\includegraphics[scale=0.41]{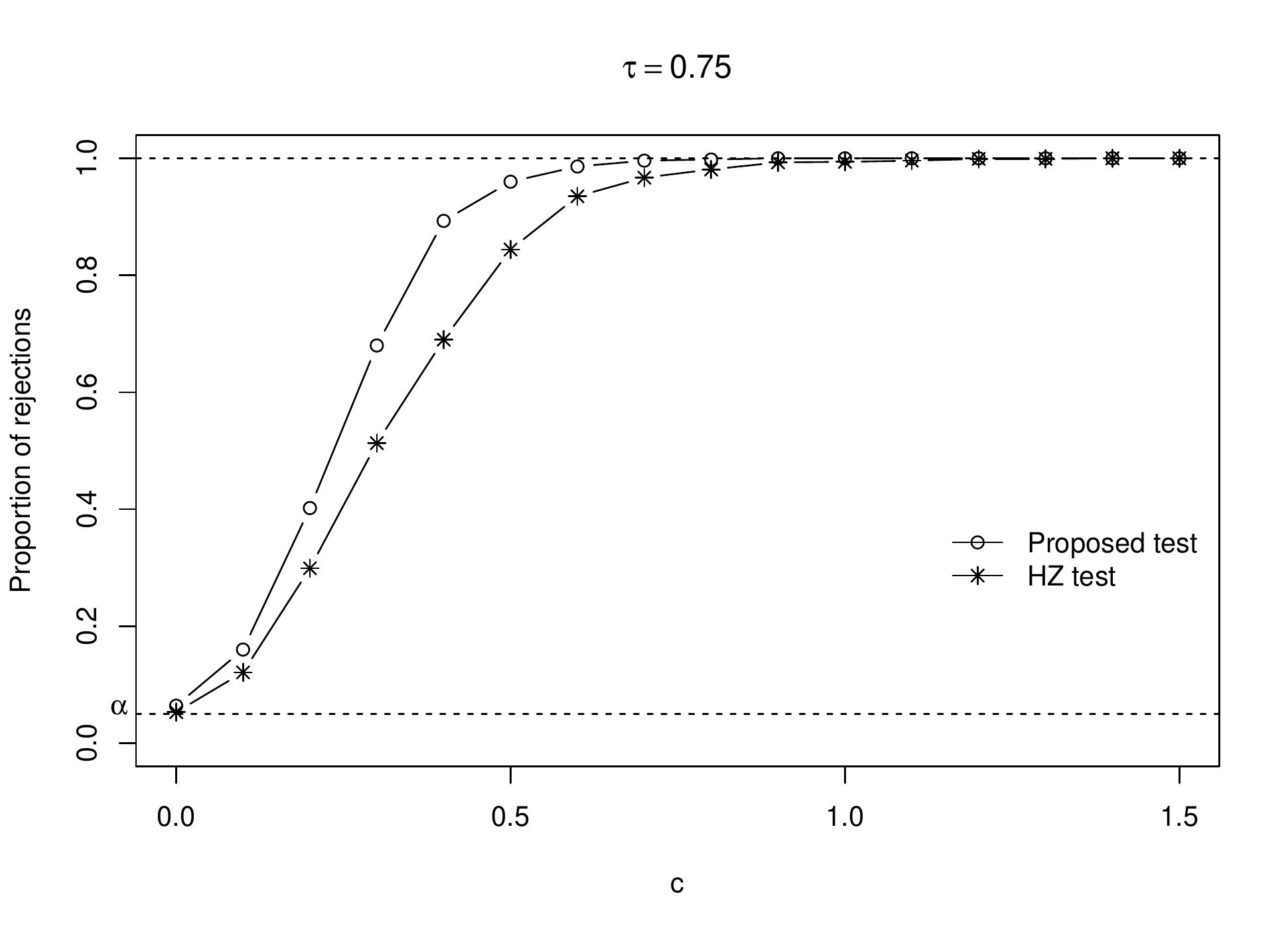}
\end{minipage}
\begin{minipage}{.49\linewidth}
\centering
\includegraphics[scale=0.41]{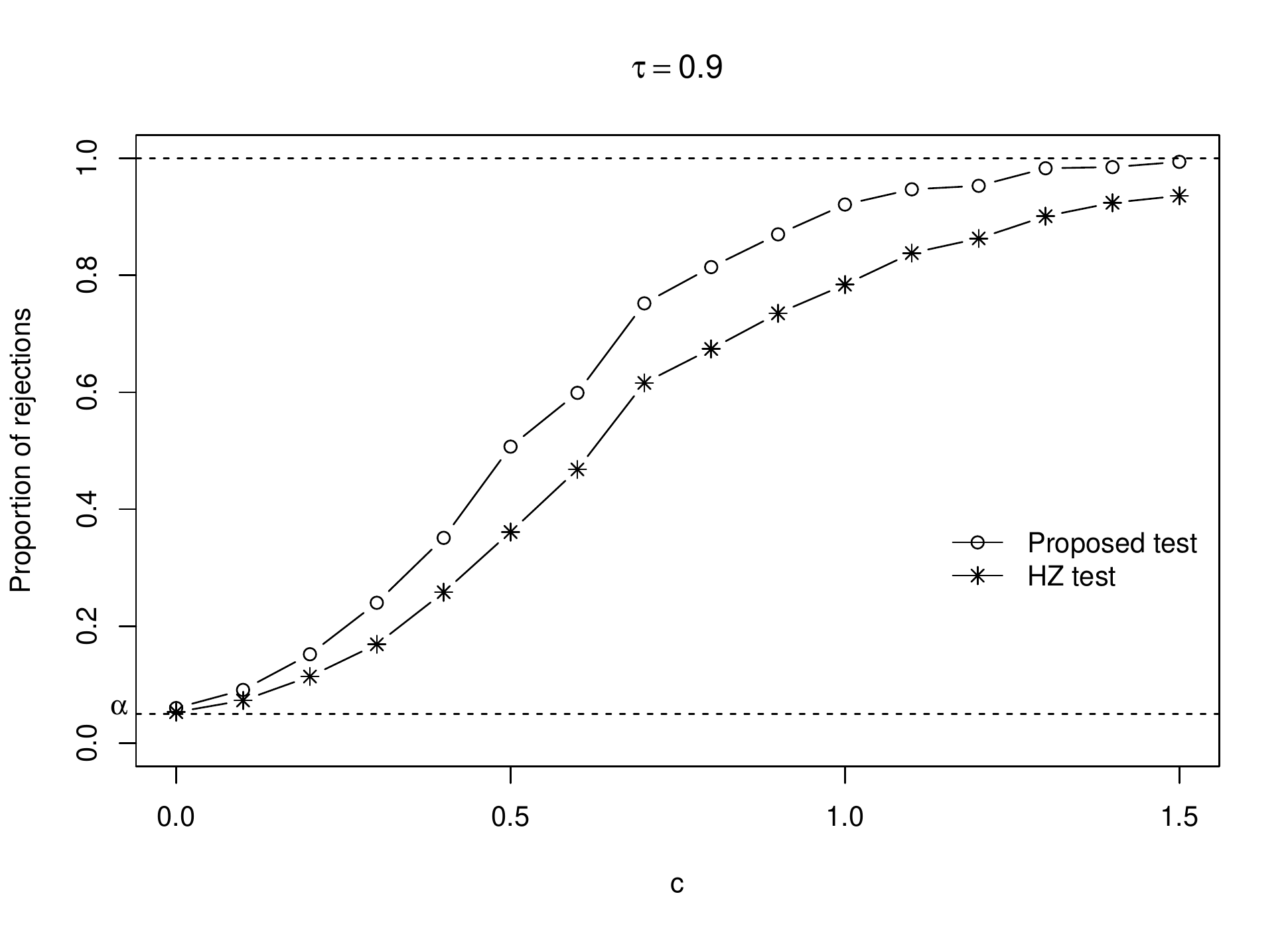}
\end{minipage}
\caption{Proportion of rejections associated with our proposed lack-of-fit test (Proposed test) and the test proposed by  \citet{HyZ03} (HZ test) for Model 6 depending on the parameter $c$ and the $\tau$-quantile of interest.}
\label{figure1}
\end{figure}

Figure \ref{figure1} shows the powers of the proposed test and \citet{HyZ03}'s test as functions of the value of $c$, and with five orders of the quantile: $0.1$, $0.25$, $0.5$, $0.75$, and $0.9$. The nominal level is $\alpha=0.05$ and the sample size is fixed to $n=150$. As expected, the power increases with $c$. The new test is more powerful than \citet{HyZ03}'s test for any value of $c$ and for any of the considered quantiles. Both tests are more powerful for central quantiles (orders close to $0.5$). Symmetry around $0.5$ is not strictly satisfied, since the error distribution is not symmetric around the median, and the deviation from the null hypothesis is more complex than that given in Model 5.

We consider different deviations from the linear null hypothesis and error distributions, as Model 7.

\begin{equation*}
\mbox{Model 7:} \; \; \; Y=1+X_1+X_2+h(X)+\varepsilon,
\label{modelo7}
\end{equation*}
where $X_{1} \in \mbox{Uniform}(0,1)$, $X_{2} \in N(0,1)$; and $\varepsilon=Z-z_\tau$, with $z_\tau$ being the $\tau$-quantile of the variable $Z$; and $X_1$, $X_2$, and $Z$ are drawn independently. For the deviation $h(X)$, a quadratic function including interaction is considered, as well as a sinus, exponential, and logarithm function of the linear transformation $l(x)=1+x_1+x_2$ (see Table \ref{table5}). For the distribution of $Z$, the log-normal, chi-squared with two degrees of freedom, exponential with expectation one, and a mixture of normal distributions are considered. The mixture is obtained as a standard normal with probability $0.75$ and a normal distribution with mean $5$ and standard deviation $2$ with probability $0.25$.

The proposed test and \citet{HyZ03}'s test are applied to check the null hypothesis of linearity on $X_1$ and $X_2$ with nominal level $\alpha=0.05$. Results for the proportions of rejections are given in Table \ref{table5}. For each deviation and each error distribution, the proposed test is more powerful than \citet{HyZ03}'s.

\renewcommand {\arraystretch}{1.3}
\renewcommand {\tabcolsep}{10pt}
\begin{table}
\begin{center}
\scalebox{0.65}[1]{
\begin{tabular}{|c| c c|c c | c c | c c | c c|}
\cline{4-11}
\multicolumn{3}{c|}{}&\multicolumn{2}{c|}{$Z\in e^{N(0,1)}$}
&\multicolumn{2}{c|}{$Z\in\chi^{2}_{2}$}
&\multicolumn{2}{c|}{$Z\in Exp(1)$}
&\multicolumn{2}{c|}{$Z\in \mbox{Mixture}$}\\
\cline{4-11}
\multicolumn{3}{c|}{} & Proposed & HZ & Proposed & HZ & Proposed & HZ & Proposed & HZ \\
\hline
\hline
$h(x)=\frac{1}{3}(x_{1}^2+x_{2}^2+x_{1}x_{2})$
	& $n=50$ & $\tau=0.25$ & 0.373 & 0.162 & 0.199 & 0.097 & 0.448 & 0.184 & 0.135 & 0.083 \\
    &        & $\tau=0.5 $ & 0.577 & 0.364 & 0.345 & 0.208 & 0.696 & 0.435 & 0.287 & 0.175 \\
    &        & $\tau=0.75$ & 0.309 & 0.217 & 0.200 & 0.150 & 0.490 & 0.365 & 0.074 & 0.068 \\
\cline{2-11}
	& $n=150$ & $\tau=0.25$ & 0.994 & 0.910 & 0.934 & 0.705 & 1.000 & 0.962 & 0.702 & 0.386 \\
    &         & $\tau=0.5 $ & 0.981 & 0.899 & 0.849 & 0.619 & 0.999 & 0.976 & 0.829 & 0.617 \\
    &         & $\tau=0.75$ & 0.773 & 0.579 & 0.516 & 0.361 & 0.952 & 0.831 & 0.138 & 0.112\\
\hline
$h(x)=5 \; sin(0.6\; \pi \; l(x))$
	& $n=50 $ & $\tau=0.25$ & 0.443 & 0.409 & 0.425 & 0.414 & 0.461 & 0.429 & 0.381 & 0.356 \\
	&         & $\tau=0.5 $ & 0.562 & 0.321 & 0.458 & 0.270 & 0.607 & 0.353 & 0.390 & 0.238 \\
	&         & $\tau=0.75$ & 0.124 & 0.066 & 0.106 & 0.061 & 0.157 & 0.081 & 0.103 & 0.053 \\
\cline{2-11}
	& $n=150$ & $\tau=0.25$ & 1.000 & 0.996 & 1.000 & 0.990 & 1.000 & 0.999 & 0.998 & 0.985 \\
	&         & $\tau=0.5 $ & 1.000 & 0.997 & 1.000 & 0.998 & 1.000 & 1.000 & 1.000 & 0.957 \\
	&         & $\tau=0.75$ & 0.865 & 0.419 & 0.811 & 0.380 & 0.982 & 0.637 & 0.586 & 0.228 \\
\hline
$h(x)=8 \; exp(-0.5 \; l(x))$
	& $n=50 $ & $\tau=0.25$ & 0.190 & 0.113 & 0.154 & 0.112 & 0.169 & 0.135 & 0.133 & 0.109 \\
	&         & $\tau=0.5 $ & 0.411 & 0.251 & 0.254 & 0.167 & 0.483 & 0.268 & 0.225 & 0.161 \\
	&         & $\tau=0.75$ & 0.244 & 0.145 & 0.164 & 0.097 & 0.382 & 0.251 & 0.102 & 0.089 \\
\cline{2-11}
	& $n=150$ & $\tau=0.25$ & 0.917 & 0.498 & 0.788 & 0.378 & 0.963 & 0.577 & 0.533 & 0.281 \\
	&         & $\tau=0.5 $ & 0.980 & 0.747 & 0.797 & 0.455 & 0.998 & 0.868 & 0.759 & 0.458 \\
	&         & $\tau=0.75$ & 0.700 & 0.450 & 0.493 & 0.325 & 0.955 & 0.744 & 0.216 & 0.137 \\
\hline
$h(x)=6 \; \log |l(x)|$
	& $n=50 $ & $\tau=0.25$ & 0.820 & 0.627 & 0.736 & 0.570 & 0.874 & 0.678 & 0.622 & 0.483 \\
    &         & $\tau=0.5 $ & 0.396 & 0.306 & 0.291 & 0.200 & 0.561 & 0.398 & 0.227 & 0.183 \\
    &         & $\tau=0.75$ & 0.090 & 0.094 & 0.098 & 0.086 & 0.122 & 0.104 & 0.068 & 0.074 \\
\cline{2-11}
	& $n=150$ & $\tau=0.25$ & 1.000 & 0.998 & 0.999 & 0.987 & 1.000 & 1.000 & 0.997 & 0.973 \\
	&         & $\tau=0.5 $ & 0.897 & 0.757 & 0.751 & 0.558 & 0.971 & 0.875 & 0.660 & 0.471 \\
	&         & $\tau=0.75$ & 0.166 & 0.171 & 0.167 & 0.166 & 0.297 & 0.196 & 0.112 & 0.138 \\
\hline
\end{tabular}}
\caption{Proportions of rejections associated with our proposed lack-of-fit test (Proposed) and to the test proposed by \citet{HyZ03} (HZ) for Model 7.}

\label{table5}
\end{center}
\end{table}

Our main purpose in proposing a new lack-of-fit test was to overcome the curse of dimensionality. Thus, the new test should show an acceptable power for increasing dimensionality of the covariate. To check this, we simulate values of the following median regression model:
\begin{equation*}
\mbox{Model 8:} \; \; \; Y=1+X_{1}+X_{2}+\frac{1}{3}\biggl(X_{1}^{2}+X_{1}X_{2}+X_{2}^{2}\biggr)+\varepsilon,
\end{equation*}
where our goal is to realize the following lack-of-fit test:
\begin{equation*}
\begin{cases}
& H_0: Y=\theta_0+\theta_1X_1+\theta_2X_2+\varepsilon \\
& H_a: Y=g(X_1,X_2,X_{2+1},\ldots,X_{2+t})+\varepsilon,
\end {cases}
\end{equation*}
where $X_{i} \in \mbox{Uniform}(0,1)$ if $i$ is odd, and $X_{i} \in N(0,1)$ if $i$ is even; the error is drawn from the centered log-normal distribution, i.e., $\varepsilon=e^{Z}-1$ where $Z \in N(0,1)$; $g$ is any smooth (nonparametric) function of the covariates; and $t$ represents the number of additional covariates in the alternative, and so is the additional dimension where the test is looking for deviations from the null. It would be expected that increased value of $t$ implies decreased power of the test.

Table \ref{table6} shows the proportions of rejections associated with the new test and \citet{HyZ03}'s test, for different values of the additional dimension, $t$. Both tests suffer a loss of power due to the increase of the dimension, as expected. Nonetheless, the loss of power is more pronounced for the test proposed by \citet{HyZ03}. For example, from dimension $t=6$ the proportion of rejections associated with their test is near to the significance level, whereas our proposed test preserves noticeable power, even for very high dimensions.

Note that, for very high dimensions, \citet{HyZ03}'s test statistic is almost degenerate, because for any observation of the covariate, $X_i$, the indicators $I(X_j\leq X_i)$, involved in the computation of their test process at $X_{i}$, will be zero for most of the other observations $X_j$, when the dimension of the covariates $X_i$ and $X_j$ is large. Thus, the test is unable to make a reasonable number of evaluations to check the model, and its power is consequently destroyed, as observed in Table 7 for $t>10$. On the other hand, our proposed method is able to make comparisons even for large dimensions of the covariate, because the indicators are calculated with unidimensional projections of the covariate. We conclude that the proposed method constitutes a necessary modification of \citet{HyZ03} when the dimension of the covariate is large.

\begin{table}
\begin{center}
\scalebox{0.65}[1]{
\begin{tabular}{|c c|c c c|c c c|}
\cline{3-8}
\multicolumn{2}{c|}{}&\multicolumn{3}{c|}{Proposed test}&\multicolumn{3}{c|}{HZ test}\\
\cline{3-8}
\multicolumn{2}{c|}{}&$\alpha=0.10$&$\alpha=0.05$&$\alpha=0.01$&$\alpha=0.10$&$\alpha=0.05$&$\alpha=0.01$\\
\hline
\hline
$t=0$ &$n=25$  & 0.252 & 0.154 & 0.057 & 0.225 & 0.135 & 0.035 \\
      &$n=50$  & 0.675 & 0.564 & 0.361 & 0.487 & 0.357 & 0.163 \\
      &$n=100$ & 0.961 & 0.918 & 0.776 & 0.822 & 0.725 & 0.460 \\
      &$n=150$ & 0.993 & 0.983 & 0.943 & 0.949 & 0.903 & 0.751 \\
      &$n=200$ & 0.999 & 0.998 & 0.990 & 0.982 & 0.965 & 0.897 \\
\hline
$t=2$ & $n=25$  & 0.177 & 0.100 & 0.029 & 0.143 & 0.080 & 0.021 \\
      & $n=50$  & 0.507 & 0.391 & 0.186 & 0.215 & 0.117 & 0.040 \\
      & $n=100$ & 0.868 & 0.813 & 0.638 & 0.349 & 0.228 & 0.077 \\
      & $n=150$ & 0.978 & 0.957 & 0.869 & 0.506 & 0.355 & 0.163 \\
      & $n=200$ & 0.997 & 0.993 & 0.975 & 0.636 & 0.498 & 0.263 \\
\hline
$t=6$ & $n=25$  & 0.133 & 0.055 & 0.010 & 0.054 & 0.018 & 0.004 \\
      & $n=50$  & 0.345 & 0.244 & 0.097 & 0.098 & 0.051 & 0.010 \\
      & $n=100$ & 0.797 & 0.696 & 0.501 & 0.097 & 0.056 & 0.021 \\
      & $n=150$ & 0.935 & 0.901 & 0.768 & 0.151 & 0.083 & 0.027 \\
      & $n=200$ & 0.992 & 0.978 & 0.929 & 0.177 & 0.089 & 0.029 \\
\hline
$t=10$ & $n=25$  & 0.120 & 0.057 & 0.011 & 0.066 & 0.018 & 0.005 \\
       & $n=50$  & 0.267 & 0.161 & 0.056 & 0.043 & 0.028 & 0.004 \\
       & $n=100$ & 0.659 & 0.562 & 0.366 & 0.052 & 0.025 & 0.003 \\
       & $n=150$ & 0.884 & 0.830 & 0.672 & 0.071 & 0.036 & 0.006 \\
       & $n=200$ & 0.966 & 0.946 & 0.887 & 0.085 & 0.040 & 0.008 \\
\hline
$t=20$ & $n=25$  & 0.094 & 0.042 & 0.010 & 0.065 & 0.023 & 0.010 \\
       & $n=50$  & 0.174 & 0.098 & 0.019 & 0.055 & 0.024 & 0.007 \\
       & $n=100$ & 0.520 & 0.398 & 0.235 & 0.054 & 0.028 & 0.005 \\
       & $n=150$ & 0.800 & 0.707 & 0.525 & 0.000 & 0.004 & 0.003 \\
       & $n=200$ & 0.918 & 0.876 & 0.748 & 0.050 & 0.033 & 0.008 \\
\hline
$t=50$ & $n=25$  & 0.074 & 0.044 & 0.005 & 0.050 & 0.026 & 0.007 \\
       & $n=50$  & 0.111 & 0.059 & 0.014 & 0.074 & 0.036 & 0.009 \\
       & $n=100$ & 0.237 & 0.149 & 0.041 & 0.068 & 0.034 & 0.007 \\
       & $n=150$ & 0.492 & 0.374 & 0.188 & 0.001 & 0.005 & 0.005 \\
       & $n=200$ & 0.686 & 0.600 & 0.438 & 0.063 & 0.024 & 0.009 \\
\hline
\end{tabular}}
\caption{Proportions of rejections associated with our proposed lack-of-fit test (Proposed test) and the test proposed by \citet{HyZ03} (HZ test) for Model 8.}
\label{table6}
\end{center}
\end{table}

\section{Application to real data} \label{sec3}

The proposed method is applied to real data from the evolution of the Gross Domestic Product (GDP) in several countries. GDP is an economic indicator that reflects the monetary value of the goods and final services produced by an economy in a certain period and it is used as a measure of the material well-being of a society. Different median regression models have been proposed to explain the annual growth rate of the Per Capita GDP in terms of a number of explanatory variables, including the initial Per Capita GDP and diverse economic and social indicators.

We focus on the model of \citet{KyM99}, based on the available information included in \citet{barro1}. A complete study of this economic model is given by \citet{barro2}. The aim of \citet{KyM99} was to check the combined effect of the different explanatory variables on the response in a quantile regression model. Here we test the specification of the quantile regression model itself.

The dataset we use is available in the {\bf R} package {\bf quantreg}, {\bf barro} (\url{http://cran.at.r-project.org/web/packages/quantreg/}). This data set contains measurements associated with 71 countries during the period 1965-1975 and 90 countries during the period 1975-1985, yielding a total sample size of $n=161$ countries.

The explanatory variables used to explain the median of the annual growth of the Per Capita GDP (the response variable, $Y$) can be split in two groups as given below. More details about these variables and their role in the model for GDP can be found in \citet{barro2}.
\begin{description}
	\item [State variables:] These variables reflect characteristics of the different countries that cannot be directly decided by political or social agents. They are measures of the steady-state position of the country, such as human capital, education or health. \citet{KyM99} consider the following variables in this group:
\begin{flushleft}
\noindent
\hspace*{2cm}$X_{1}:=$ log(Initial Per Capita GDP)\\
\hspace*{2cm}$X_{2}:=$ Male Secondary Education\\
\hspace*{2cm}$X_{3}:=$ Female Secondary Education\\
\hspace*{2cm}$X_{4}:=$ Female Higher Education\\
\hspace*{2cm}$X_{5}:=$ Male Higher Education\\
\hspace*{2cm}$X_{6}:=$ Life Expectancy\\
\hspace*{2cm}$X_{7}:=$ Human Capital\\
\end{flushleft}
	\item [Control and environmental variables:] These variables are direct consequences of decisions made by government or private agents. The variables included in this group are
\begin{flushleft}
\hspace*{2cm}$X_{8}:=$ Education/GDP\\
\hspace*{2cm}$X_{9}:=$ Investment/GDP\\
\hspace*{2cm}$X_{10}:=$ Public Consumption/GDP\\
\hspace*{2cm}$X_{11}:=$ Black Market Premium\\
\hspace*{2cm}$X_{12}:=$ Political Instability\\
\hspace*{2cm}$X_{13}:=$ Growth Rate Terms Trade
\end{flushleft}	
\end{description}

We apply the AIC criterion proposed by \citet{HyT90} to variable selection among the thirteen explanatory variables for the quantile regression model. We will consider only those variables that show as relevant for the response. Based on this criterion, we propose a model that includes the variables $X_i$ with $i \in \mathcal{I}_{1}=\{1,2,6,7,9,10,11,12,13\}$.

We apply the proposed lack-of-fit test in four different testing problems:
\begin{equation*}
\begin{split}
& \quad\text {Problem 1 \;}
\left\{
\begin{aligned}
& H_0: Y=\theta_0+\sum_{i=1}^{13}X_i'\theta_i+\varepsilon_{1} \\
& H_a: Y=g(X_1,X_2,\ldots,X_{13})+\varepsilon_{1}
\end {aligned}
\right.\\
& \\
& \quad\text {Problem 2 \;}
\left\{
\begin{aligned}
& H_0: Y=\theta_0+\sum_{i \in \mathcal{I}_{1}}X_i'\theta_i+\varepsilon_{2} \\
& H_a: Y=g(X_i, i \in \mathcal{I}_{1})+\varepsilon_{2}\end {aligned}
\right. \\
& \\
& \quad\text {Problem 3 \;}
\left\{
\begin{aligned}
& H_0: Y=\theta_0+\sum_{i \in \mathcal{I}_{1}}X_i'\theta_i+\varepsilon_{3} \\
& H_a: Y=g(X_1,X_2,\ldots,X_{13})+\varepsilon_{3}\end {aligned}
\right.\\
& \\
& \quad\text {Problem 4 \;}
\left\{
\begin{aligned}
& H_0: Y=\theta_0+\sum_{i \in \mathcal{I}_{2}}X_i'\theta_i+\varepsilon_{4} \\
& H_a: Y=g(X_1,X_2,\ldots,X_{13})+\varepsilon_{4}\end {aligned}
\right.
\end {split}
\end{equation*}
where $\mathcal{I}_{2}=\{1,2,3,4,5,6,7\}$ (state variables). Problem 1 is a lack-of-fit test of the linear model versus a nonparametric alternative, including all the thirteen explanatory variables under both the null and alternative hypotheses. Problem 2 is a lack-of-fit test of the linear model versus a nonparametric alternative, including only the nine variables in the set $\mathcal{I}_{1}$. Problem 3 is the same test as Problem 2, but with an alternative in the thirteen original variables. Problem 4 is a lack-of-fit test of a linear model that only includes the state variables.

Table \ref{table7} contains the $p$-values obtained from the application of the proposed lack-of-fit test to each of the testing problems. The number of bootstrap replications was $B=500$. We would accept the null hypothesis in Problems 1, 2 and 3. In Problem 3, the model under the null is the simplest, while the model under the alternative is the most complex. Despite this, the $p$-value is quite large, so we can conclude that the simple model with the nine explanatory variables in the set $\mathcal{I}_{1}$ is correct, and there is no significant deviation from this model arising from any (smooth) function of the thirteen possible explanatory variables.

On the other hand, the null hypothesis is rejected for Problem 4. Thus, a model that only includes the state variables is insufficient to explain the evolution of the GDP, that is, some of the control or environmental variables are necessary.

In summary, our proposed test confirms the validity of the model proposed by \citet{KyM99}. In addition, from the outcome for Problem 3, it would be sufficient to consider a model with nine explanatory variables to explain the growth rate of the Per Capita GDP. 

\renewcommand {\arraystretch}{1.3}
\renewcommand {\tabcolsep}{10pt}
\begin{table}
\begin{center}
\scalebox{0.65}[1]{
\begin{tabular}{|r|c|c|c|c|}
\cline{2-5}
\multicolumn{1}{c|}{}&\multicolumn{1}{c|}{Problem 1}&\multicolumn{1}{c|}{Problem 2}&\multicolumn{1}{c|}{Problem 3}&
\multicolumn{1}{c|}{Problem 4}\\
\hline
$p$-values & 0.194 & 0.458 & 0.440 &  0.002 \\
\hline
\end{tabular}}
\caption{$p$-values obtained by the proposed  lack-of-fit test for Problems 1, 2, 3 and 4.}
\label{table7}
\end{center}
\end{table}

\section{Concluding remarks and extensions}

We proposed a new lack-of-fit test for quantile regression models, together with a bootstrap mechanism to approximate the critical values. The bootstrap approximation does not need to estimate the conditional sparsity, and was shown to work well in homoscedastic and heteroscedastic error distributions and with high-dimensional covariates.

The proposed test is generally more powerful than its natural competitors, and particularly more powerful in the case of a high-dimensional covariate.

The proposed test was applied to a real data situation, where it was useful to validate well-known models in the economic literature, that describe the evolution of the GDP in terms of a number of explanatory variables.

The proposed method can be generalized to test models involving quantiles of different orders. The most treated model in the literature is the multiple quantile linear model, where it is assumed that the quantile regression function is linear for a subset of orders $\tau\in{\mathcal T}\subset[0,1]$, 
$$g_\tau(x)=x'\theta(\tau),$$
with coefficients $\theta(\tau)$ depending on the order, $\tau$, of the quantile. The coefficients $\theta(\tau)$ allow consideration of a different effect of the covariates depending on the order of the quantile. See \citet{EyG14} for a lack-of-fit test of multiple quantile linear models, or \citet{EyV10} for a test of multiple quantile models with time series. Our proposed method can be generalized to test multiple quantile models in a general framework of parametric (possibly nonlinear) quantile regression with heteroscedasticity and without estimating unknown quantities. To this end, one would consider a process depending on $(\beta,u)$, as well as on $\tau$. We restricted to the case of testing a single quantile to focus on the performance of the test for high-dimensional covariates and other important features of the testing problem. Extension to multiple quantile testing was left to future research. Similarly, extensions of the proposed method to time series are possible using the results in \citet{EyV10}. These possible extensions show that the concept of projecting the covariate, given by \citet{Esc06} to overcome the curse of dimensionality, combined with the bootstrap methodology introduced by \citet{FHH11}, provide a promising strategy for checking quantile regression models.

\section*{Acknowledgements}

The authors are grateful for constructive comments from the associate editor and three reviewers which helped to improve the paper. They also acknowledge the support of Project MTM2008-03010, from the Spanish Ministry of Science and Innovation and IAP network StUDyS, from Belgian Science Policy. Work of M. Conde-Amboage has been supported by FPU grant AP2012-5047 from the Spanish Ministry of Education.


\end{document}